\documentclass[runningheads]{llncs}
\usepackage{graphicx}
\usepackage{amsmath}
\usepackage{cases}
\usepackage[table]{xcolor}
\usepackage[colorinlistoftodos,prependcaption,textsize=tiny]{todonotes}
\usepackage{gensymb}
\usepackage{textgreek}
\usepackage[caption=false]{subfig}
\usepackage{multirow}

\begin{document}



\title{Sonification as a Reliable Alternative to Conventional Visual Surgical Navigation}

\titlerunning{Sonification as a Reliable Alternative ...}
%
%
\author{Sasan Matinfar*\inst{1,4} \and Mehrdad Salehi\inst{1} \and Daniel Suter\inst{2} \and Matthias Seibold\inst{1,3} \and Navid Navab\inst{5} \and Shervin Dehghani\inst{1,4} \and Florian Wanivenhaus\inst{2} \and Philipp F\"urnstahl\inst{2,3} \and Mazda Farshad\inst{2} \and Nassir Navab\inst{1}
}
\authorrunning{S. Matinfar, M. Salehi et al.}
%
\institute{Computer Aided Medical Procedures, Technical University of Munich, Germany \and 
Department of Orthopaedics, Balgrist University Hospital, University of Zurich, Switzerland \and
Research in Orthopedic Computer Science Group, Balgrist University Hospital, University of Zurich, Zurich, Switzerland \and
Nuklearmedizin rechts der Isar, Technical University of Munich, Germany \and
Topological Media Lab, Concordia University, Montreal, Canada}

\maketitle            
%
\begin{abstract}
Despite the undeniable advantages of image-guided surgical assistance systems in terms of accuracy, such systems have not yet fully met surgeons’ needs or expectations regarding usability, time efficiency, and their integration into the surgical workflow. On the other hand, perceptual studies have shown that presenting independent but causally correlated information via multimodal feedback involving different sensory modalities can improve task performance. This article investigates an alternative method for computer-assisted surgical navigation, introduces a novel sonification methodology for navigated pedicle screw placement, and discusses advanced solutions based on multisensory feedback. The proposed method comprises a novel sonification solution for alignment tasks in four degrees of freedom based on frequency modulation (FM) synthesis. We compared the resulting accuracy and execution time of the proposed sonification method with visual navigation, which is currently considered the state of the art. We conducted a phantom study in which 17 surgeons executed the pedicle screw placement task in the lumbar spine, guided by either the proposed sonification-based or the traditional visual navigation method. The results demonstrated that the proposed method is as accurate as the state of the art while decreasing the surgeon's need to focus on visual navigation displays instead of the natural focus on surgical tools and targeted anatomy during task execution.

\keywords{Sonification \and Pedicle Screw Placement \and Navigation \and Multisensory Processing \and Auditory Feedback \and Computer Assisted Intervention}
\end{abstract}

\section{Introduction}
\label{Introduction}
Computer-assisted navigation systems provide surgeons with rich and complex multimodal data, enhancing intraoperative diagnosis, decision making, and surgical maneuvers. Despite the high reliability of such systems, they have not yet been fully integrated into the surgical workflow. The dominant way of conveying information in current navigation systems is based on visual displays, a method that assists the surgeon only via the unisensory perceptual channel of vision. This may be explained because that we are biologically trained to localize objects, including their semantic meaning, visually, based on a Cartesian grid in a static form. However, in a dynamic interaction with a navigation system, occurring over time, objects' qualities are constantly transforming into new states. This challenges the surgeon's cognition, creating complications, especially in a high-intensity environment such as an operating room. A challenge related to hand--eye coordination is that the surgeon's visual attention has to diverge between navigation displays and the actual operation area, including the surgical tools, targeted anatomy, and the surrounding critical structures. Such complications have not been completely resolved even in more recent augmented reality (AR)-based systems, when overloading multiple virtual visual cues on the display may lead to change or inattentional blindness~\cite{inattentional1,inattentional2}. 

In cognitive psychological research, it has been shown that multisensory integration facilitates information processing. Multisensory integration, that is, the combination of multiple independent but causally correlated information sources from different senses, including auditory, visual, and haptic, improves performance on a wide range of tasks ~\cite{cognitive1,cognitive2}. Research in computer-assisted surgery has not yet fully taken advantage of multisensory feedback and there are unanswered questions in this regard. In this article, we have highlighted the importance of alternative perceptual modalities for navigated surgery, investigating potential solutions and discussing the future picture of surgical navigation systems. The human auditory perception, as opposed to the visual, is not tied to a spatialized atemporal Cartesian grid. Therefore, sonic qualities such as texture, timbre, and rhythm, which unfold over time, are more efficient and more convenient to embody temporal aspects of objects' qualities. The idea of using sound as a source of information has been well founded in sonification research, which is often defined as the systematic transformation of data relations into perceived relations in an acoustic signal to facilitate communication or interpretation that is reproducible~\cite{kramer,sonhand,sonificationtaxonomy}. The auditory channel as an alternative perceptual modality to visual feedback has proven to be beneficial in different domains, such as process monitoring, data exploration, and navigation~\cite{sonificationhandbook,SID}. Incorporating navigation data into multiple alternative channels will unload a single modality, creating new possibilities for presenting interaction data with computer systems more intuitively. The challenge of sonification design for surgical navigation is to incorporate the complex dimensionality of the application scenario into an integrated audio stream, that meets the clinician's expectations in terms of reliability, usability, and time efficiency. 

We hypothesize that a multisensory-based navigation system improves the surgeon's perception in highly precise interventional tasks. This article, as the first step toward multisensory navigation, introduces a novel standalone sonification methodology for the pedicle screw placement task in lumbar spine surgery. To demonstrate the feasibility of the solution, we evaluated the method in a phantom study with 17 orthopedic surgeons in terms of effectiveness, usability, and learnability in comparison with conventional 3D visual navigation as an established method and state of the art with respect to accuracy. Despite the fact that the surgeons have more experience using visual feedback, the study results confirmed the reliability of sonification for surgical navigation tasks and demonstrated the potential behind the core idea of this research.

\section{Clinical Motivation}
\label{clinicalMotivation}
Severe pathological conditions of the spine, including deformity, trauma, degenerative disc disease, and spondylolisthesis, can be treated using the established orthopedic surgical technique called spinal fusion or spondylodesis~\cite{spinal-fusion-usa,systematic-spinal-fusion}. Spinal fusion implants, which consist of specialized screws that are driven into the pedicles of the respective vertebrae, are used to achieve a fusion between two or more spine segments, thereby immobilizing the respective region and absorbing biomechanical forces. In modern approaches, the surgeon prepares a guiding hole for the smooth insertion of screws, using a surgical awl or by drilling K-wires. To determine the central position of the guiding hole within the pedicle, the surgeon uses bony landmarks for orientation~\cite{screw1,screw2}. Optimal positioning is crucial for avoiding screw perforation, which can cause serious injury to the spinal cord and its surrounding nerves and vessels. Hence, accurate pedicle screw placement is essential for a surgical outcome, and success depends on the experience and anatomical understanding of the surgeon, especially in severe cases such as scoliosis, kyphosis, or congenital anomalies, where the chance of perforation is even greater~\cite{guideline-screw}. 

There are three main techniques for pedicle screw placement: freehand, fluoroscopy guidance, and stereotactic navigation~\cite{systematic-pedicle,three}. The misplacement rate, that is, the rate of screws perforating the pedicle cortex to any degree, in the freehand technique ranges from 5\% to 41\% in the lumbar spine and from 3\% to 55\% in the thoracic spine~\cite{systematic-pedicle}. The high rates of misplaced screws in the freehand approach, various pedicle morphology, and different sizes of the vertebral body motivate computer-assisted systems to improve surgical accuracy~\cite{systematic-pedicle}. However, there exists some level of disagreement about the necessity of accuracy in pedicle screw placement~\cite{accuracy-review}. A careful analysis of related studies~\cite{screw1,screw2,guideline-screw,systematic-pedicle,accuracy-review,accuracy-safety} shows that accuracy and safety are dependent on several factors, such as the vertebrae level in question, the definitions of thresholds and safety zones, whether the pedicle cortex has been perforated or not, the applied technique, and the availability of the dataset for comparison studies. There have been studies~\cite{accuracy-safety,freehandpos1} that considered the freehand technique an accurate and safe technique for pedicle screw placement, and many surgeons believe that even when performed slightly inaccurately, such imprecise pedicle screw placement is asymptomatic. However, even those asymptomatic cases can cause implant instabilities, prevent smooth fusion, or expedite adjacent-level degeneration~\cite{accuracy-review,accurate}. Using conventional fluoroscopy has not entirely solved the problem, as the misplacement rate has been reported as 31.9\%~\cite{accuracy-review} and even higher in more challenging cases~\cite{accuracy-safety}.

Conversely, computer-assisted systems for pedicle screw navigation have been shown to be more accurate, with reduced complications~\cite{navigation1,navigation2,navigation3,navigation4,navigation5}. Intraoperative image-guided navigation has evolved in recent years as established approaches such as 2D and 3D fluoroscopic navigation have increased the rate of successful placements respectively to 84.3\% and 95.5\%, respectively~\cite{accuracy-review}. Furthermore, computer-assisted navigation avoids the use of intraoperative imaging, which reduces the dose of radiation required by conventional fluoroscopy~\cite{radiation1,radiation2,radiation3}. However, while the 3D fluoroscopic navigation system demonstrates the most accurate current solution for pedicle screw placement and is accepted as a standard method according to different \textit{in vivo} studies~\cite{systematic-pedicle,accuracy-review}, the adoption of such technologies in surgical workflow has been slow, requiring further system improvements~\cite{navigationdis1,navigationdis2}. In a worldwide survey on the use of navigation in spine surgery, conducted by H\"artl et al.~\cite{navigationdis1}, although 80\% of 677 participants acknowledged the use of navigation systems, they concluded current systems do not meet surgeons’ expectations in terms of usability, time efficiency, and integration into the surgical workflow. Participants complained about the complexity of use and the disruption of the surgical workflow as major factors. Additionally, they considered time-consuming training to be a prerequisite factor to support the integration of such systems, and Ryang and colleagues~\cite{learningcas} supported this in their study. Current navigation systems predominantly provide surgeons with information through visual displays, increasing the surgeons' cognitive load and complicating hand--eye coordination. Unnaturally, surgeons need to divide their focus of attention between the operation site and navigation displays~\cite{navigationdis3}, or their field of view becomes cluttered with multiple holographic cues visualized on head-mounted displays. Visual distraction is problematic for surgeons, considering they need to perceive and process complex structures of navigation data at the highest level of precision in the intensive and stressful situation of a surgical environment~\cite{inattentional1,inattentional2}.

\section{Related Studies}
Among surgical navigation systems, we focus on AR-based solutions, which include sonification as one of its emerging branches. AR has been shown to be beneficial for surgical applications~\cite{AR1,AR2}, in particular for orthopedic surgery~\cite{AR-orthopedic}. AR technology has the advantage of superimposing preoperative planning with intraoperative anatomy, which, in the case of visual-centric AR, provides surgical navigation information in the surgeon's field of view. Previous studies have proposed a body of AR-based navigation solutions for pedicle screw placement~\cite{AR-pedicle1,AR-pedicle2,AR-pedicle3,AR-pedicle4}. Similar approaches based on tool-mounted mobile devices have been used to provide information in the line of sight of a surgeon~\cite{toolmounte1,toolmounte2,toolmounte3,toolmounte4}. However, all these approaches have utilized visual feedback as the singular feedback modality. As discussed in Sections~\ref{Introduction} and~\ref{clinicalMotivation}, the inherent limitations, such as change or inattentional blindness~\cite{inattentional1,inattentional2} are the motivation behind the research reported here.

Sonification for navigation purposes was initially been designed as a natural application for people with visual impairment~\cite{blind1,blind2,sonificationnavigation1,sonificationnavigation2,sonificationnavigation3,sonificationnavigation4} and has been expanded to more general applications~\cite{sonificationnavigation5,sonificationnavigation6,sonificationnavigation7}. Sonification of one-dimensional data using primitive sound synthesis methods, such as in heart--lung machines, has already been integrated into surgical procedures. Such basic sonification methods do not extrapolate well to more complex multidimensional scenarios, as they lack consideration of psychoacoustics and sound design in their configuration. To address this problem, sonification methods~\cite{sst1,sst2,sst3} have been proposed with more focus on usability and clinical integration, using more flexible and creative sound designs; however, these approaches are unsuitable for presenting precise navigation data. 

Sonification methodologies for medical applications have mostly focused on image-guided navigation scenarios. Black et al.~\cite{sonnavmed1}, in a review paper, named three primary motivations for sonification of surgical navigation: (1) increasing awareness of structures surrounding the tracked instrument, (2) reducing attention to the screen or increasing attention to the patient or test phantom, (3) helping clinicians correctly interpret (multidimensional) navigation data. Wegner et al.~\cite{sonnavmed2} recommended different mapping ideas, such as 3D audio spatialization for generalized 3D surgical instrument placement. Sonification in the form of proximity alerts has been proposed for endoscopic cranial base surgery~\cite{sonnavmed3}, temporal bone drilling~\cite{sonnavmed4}, protecting facial nerves during otologic surgery~\cite{sonnavmed5}, guiding cochlear implantation~\cite{sonnavmed6}, and fluorescence-guided resection of gliomas~\cite{sonnavmed7}. More elaborate approaches have been introduced in~\cite{sonnavmed8,sonnavmed9,sonnavmed10,sonnavmed11} using continuous parameter-mapping sonification for surgical needle guidance in one dimension. Investigation of solutions for one-dimensional distance mapping have been undertaken by Plazak et al.~\cite{sonnavmed12}, who proposed five different mapping strategies, and Roodaki et al.~\cite{sonnavmed13}, who introduced a sonification design based on physical modeling sound synthesis that requires minimum training.

Sonification research in recent years has aimed to expand in terms of data dimensionality and degrees of freedom (DOF). Parseihian et al. ~\cite{sonificationnavigation5} investigated the efficiency of different sonification strategies in terms of rapidity and precision for a one-dimensional guidance task. Sonification of multidimensional data is challenging~\cite{sonnav1,sonnav2}, and researchers have investigated the potential of spatial sound to overcome this challenge for 2D~\cite{twod} and 3D space~\cite{threed}. Such approaches have been relatively successful when combined with visual guidance. Spatial sonification as an intuitive and natural method with a high learnability rate is suitable for orientation tasks~\cite{sonificationnavigation5,sonificationnavigation8}. However, spatial sound does not provide the precise distal and angular resolution required for precise surgical guidance tasks~\cite{spatial}. The resolution of spatial localization is $1\degree~\pm~3\degree$ along the horizontal axis in front, and becomes less toward the sides. The resolution of estimating distance is decimeters in a short distance area~\cite{psychoac}. Conversely, monaural sonification provides flexibility in design, as its efficiency is justifiable because of our inherent perceptual capability, as we can discriminate pitches in a range of 640--4000 steps~\cite{psychoac}, 120 levels of loudness~\cite{psychoac}, and 250 levels of sharpness~\cite{psychoac,jnd}. Monaural approaches are efficient regarding dimensionality and resolution; however, they introduce design challenges in terms of intuitiveness and learnability. Sonification methods are proposed for guidance in 2D~\cite{sonificationnavigation6,sonificationnavigation7} and 3D~\cite{sonificationnavigation8} spaces, providing information such as distance or orientation. These methods employed monaural sonic characteristics such as pitch, amplitude, and timbre.  

A review of the state of the art reveals a lack of research on methodologies for surgical tool guidance in two or more dimensions which would be integrable into highly sensitive application scenarios such as pedicle screw placement. In pedicle screw placement, the surgeon aligns the drill with a predefined target trajectory, which can be mathematically defined by two points, the entry and angular target points. Optimal positioning of the tool on these two points requires tool movement in four DOF. To the best of our knowledge, there is no prior study in four-DOF sonification. There are essential questions to address. For example, the approach's effectiveness and usability: to what level of precision and accuracy would sonification provide information along with an appropriate level of immediacy, and simultaneously achieve a satisfactory level of usability required for surgical situations? The majority of methods lack clinical-grade integrable sound designs, and Black et al.~\cite{sonnavmed1} described current sonification approaches as being simple. There are a limited number of studies that have compared the effect of sonification to visual feedback. Also, there is a dearth of comprehensive evaluation studies on clinical evaluations and training.

\section{Computer-assisted Auditory Navigation System}
Our approach to providing auditory navigation assistance to surgeons consists of two main components, the navigation and sonification modules. 

The navigation module comprises a workstation and an infrared optical tracking camera. The goal is to inform the surgeon intraoperatively via auditory signals of the positioning of the drill the surgeon is controlling. Prior to the operation, the trajectories of the screws are preplanned on the basis of a preoperative CT volume of the patient, and, to align the preoperative CT with the intraoperative coordinate system of the camera, a registration method is performed. Intraoperatively, using the camera, markers on the drill sleeve are tracked relative to reference markers on the patient's bed. The real-time position and orientation of the targets relative to the drill tip are sent to the workstation and used to compute error parameters (described in Subsection~\ref{errorp}). This information is, in turn, transferred to the sonification module, which generates the output sounds accordingly.

\subsection{Error Parameters}
\label{errorp}

We define the pedicle screw placement as a four-DOF alignment task between the tracked drill sleeve's tip point, $T_{tool}$, and the preoperative planned trajectory, $T_{target}$. The first two DOF correspond to the translation of $T_{tool}$'s tip point projected on the entry point plane $P_{entry}$. $P_{entry}$ is defined by taking the main direction of the planned trajectory $T_{target}$ as the plane's normal and the planned entry point to the bone as the center of the plane. Hence, both $T_{target}$ and $P_{entry}$ are both updated according to each pedicle screw's planned trajectory. The entry point errors $e_{x}$ and $e_{y}$ are defined as the distance between the center point of the $P_{entry}$ plane and the projection of the drill sleeve's tip point on $P_{entry}$. $e_{x}$ and $e_{y}$ show the entry point errors in mediolateral and caudiocranial directions, respectively.

\begin{figure}[!ht]
\includegraphics[width=\textwidth]{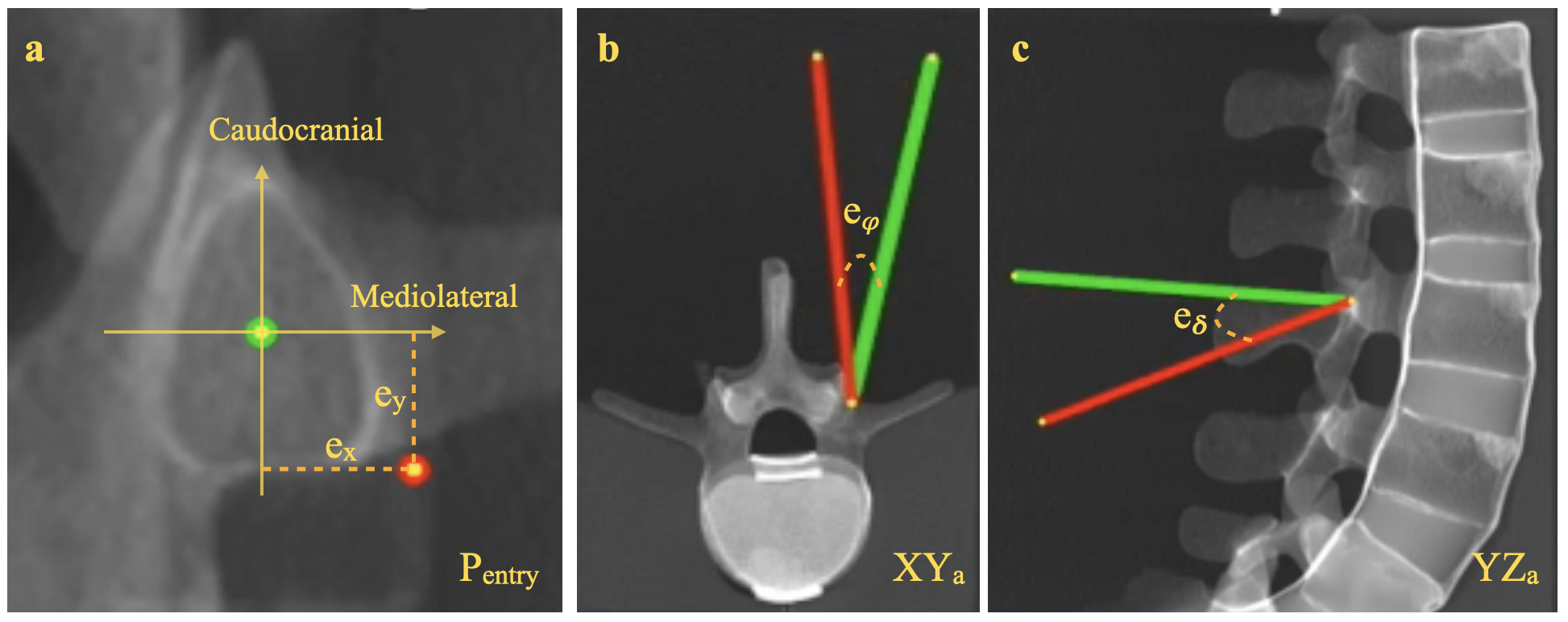}
\caption{Three cross-sectional views of the CT from the spine phantom model, including the corresponding errors ($e_{x}, e_{y}, e_{\phi}, e_{\delta}$). The target and tool are visualized in green and red, respectively. (a) Corresponds to the coronal view visualizing, $e_x$ and $e_y$ projected on the $P_{entry}$; (b) represents the axial view visualizing $e_{\phi}$; and (c) visualizes the sagittal view including $e_{\delta}$.}~\label{cross-sections}
\end{figure}

The remaining two DOF correspond to the orientation mismatch between $T_{tool}$ and $T_{target}$. This angular error is decomposed into two values, $e_{\phi}$ and $e_{\delta}$, which are Euler angle differences between the projections of the $T_{tool}$ and $T_{target}$ on the axial $XY_a$ and sagittal $YZ_a$ planes, respectively, in the anatomical coordinate system $XYZ_a$, as illustrated in Figure~\ref{cross-sections}. The orientation error on the $XZ_a$ plane is negligible because of the symmetry of the tool and the preplanned trajectory. The anatomical coordinate system stays constant throughout the execution of all pedicle screw placements.

\subsection{Four DOF Sonification Model}
\label{sonnav}
\subsubsection{Interactive Alignment Model}

The interaction model is designed with two interactive phases, namely, entry point Phase (EP), and angle phase (AP), each with two DOF. There are also two static phases, the initial phase (IP) where $T_{tool}$ has not yet entered the entry point working area ($W_{EP}$), and the final phase (FP) where $T_{tool}$ has reached $T_{target}$. First, the projections of $T_{tool}$ and $T_{target}$ on the $P_{entry}$ plane have to be aligned (EP); then, the tool orientation is aligned with $T_{target}$ (AP) while the tooltip stays in place. This implies that when the interaction is in the AP, the tooltip has already been aligned to $T_{target}$. If during the AP the tooltip deviates from $T_{target}$, the sonification will return to the sound mappings of the EP.

\begin{figure}[!ht]
\includegraphics[width=\textwidth]{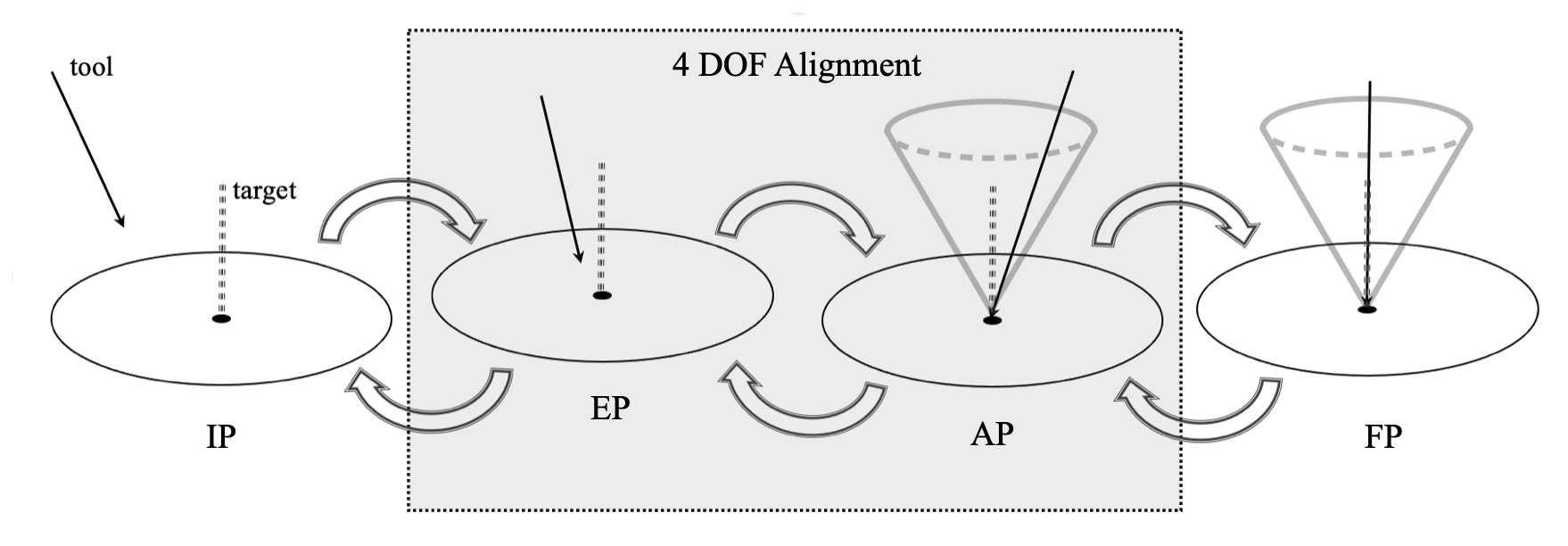}
\caption{Four DOF alignment model with four phases, initial phase (IP), entry point phase (EP), angle phase (AP), and final phase (FP). EP and AP are two interactive phases with continuous mappings, whereas IP and FP are the static phases with constant mappings.} \label{interactionalignmentmodel}
\end{figure}

The transitions between these phases and states are carried out using a threshold mechanism, with two control parameters, $d$ and $\theta$. $d$ is the 2D Euclidean distance between the projections of $T_{tool}$ and $T_{target}$ on $P_{entry}$, and $\theta$ is the 3D Euler angular distance between $T_{tool}$ and $T_{target}$. The user interaction with the sonification model starts when the tooltip enters the $W_{EP}$, which is a circle on the $P_{entry}$ plane with radius $r_{EP}$ around the target entry point. Furthermore, we define the angular working area $W_{Ang}$, which includes all $T_{tool}$s with Euler angular distance less than $\theta_{Ang}$ from the $T_{target}$; i.e., $\theta~<~\theta_{Ang}$. The alignment task is accomplished when $T_{tool}$ is aligned in all four DOF at $T_{target}$ (Fig.~\ref{interactionalignmentmodel})

\begin{figure}[!!ht]
\begin{center}
\includegraphics[width=0.8\textwidth]{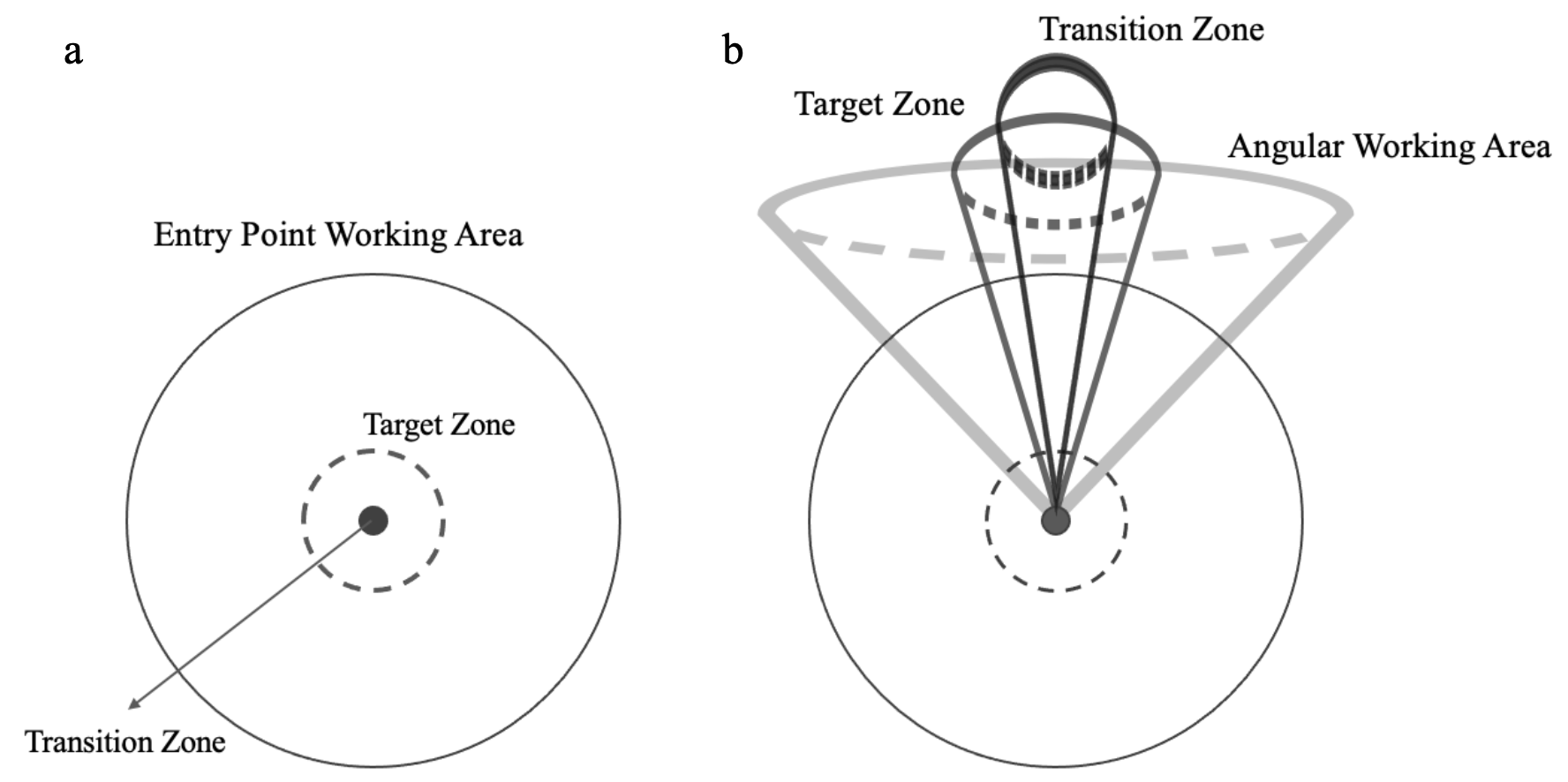}
\end{center}
\caption{Illustration of the thresholds for transition between phases. (a) the circles demonstrate thresholds for the transition between IP, EP and AP; (b) the cones represent the thresholds for transition between AP and FP.}~\label{thresholds}
\end{figure}

In interactive phases, EP and AP, we define two thresholds, namely, the target and transition zones. The transitions to a next step, that is, from EP to AP and from AP to FP, are executed only when the tool reaches inside the transition zone. When $T_{tool}$ exits the target zone, the alignment returns to a previous step, that is, from FP to AP or from AP to EP. In these cases, the user needs to reach the transition zone to be able to proceed to the next step. The threshold mechanism with the space between the target and transition zones enables us to smooth out the interaction with the system, avoiding unwanted transitions due to slight hand tremors of the surgeon or optical tracking jitter (Fig.~\ref{thresholds}).

\subsubsection{Mapping to Acoustic Features}
Sonification mapping is based on a continuous stream of pulse tones generated using the well-known FM synthesis method~\cite{fmsynth}. The input data to the sonification function are the 4D vector $(e_{x}, e_{y}, e_{\phi}, e_{\delta})$ as its parameters are described in Subsection~\ref{errorp}. These components control the fundamental frequency and the pulse rate of the pulsing stream. Depending on the alignment phase, the system controls which parameters of the input vector should be used for parameter mapping. In EP, $e_{x}$ and $e_{y}$ are used to map to a fundamental frequency and pulse rate, respectively, whereas in AP, $e_{\phi}$ and $e_{\delta}$ are used. The mapping of the pulse rate is interpolated linearly; however, exponential interpolation is used for the fundamental frequency, as the human auditory system perceives pitch in an exponential manner. 

Because both EP and AP phases use the same implementation of the synthesis function, we apply different ranges for the fundamental frequency of the FM synthesis to create higher contrast between the two alignment phases. In the IP and FP, the sonification is limited to musical major and minor chords, respectively, both pulsing at a constant rate with different values, as listed in Table~\ref{tab1}. In each interactive phase, when the $T_{tool}$ reaches the $T_{target}$ value in only one dimension, an earcon is played to facilitate the process of finding the target in the second dimension. The so-called optimum earcons consist of two sequential notes with a slight difference, depending on which target dimension has been reached. The optimum earcon for $e_{x}$ and $e_{\phi}$ is the same, whereas a slightly different earcon is used for $e_{y}$ and $e_{\delta}$. To make the transitions clear, two additional earcons were designed, consisting of eight sequential notes in ascending order for the EP to AP transition and in descending order for the AP to EP transition.

\begin{table}[ht!]
    \caption{Parameters for the FM synthesis mapping functions with the input data $e = (e_{x}, e_{y}, e_{\phi}, e_{\delta}) \in [0, 1]$ for entry point phase (EP), angle phase (AP), initial phase (IP), and final phase (FP).}\label{tab1}
    \centering
    \setlength{\tabcolsep}{5.5pt} 
    \renewcommand{\arraystretch}{1.5} 
    \begin{tabular}{c|c|c}
        phase & fundamental frequency & pulse interval\\
        \hline
        \hline
        EP & $e_{x} \xrightarrow{exp}$ [880, 1760]~Hz & $e_{y} \xrightarrow{lin} [0.35, 0.1]$~sec\\
        AP & $e_{\phi} \xrightarrow{exp}$ [110, 440]~Hz& $e_{\delta} \xrightarrow{lin} [0.35, 0.1]$~sec\\
        \hline
        IP & (123.47, 155.56, 185, 246.94)~Hz & 0.66~sec\\
        FP & (440, 523.25, 659.26, 880)~Hz & 1.5~sec\\
        
    \end{tabular}
\end{table}

\subsection{Comparison Study}
\begin{figure}[!ht]
\begin{center}
\includegraphics[width=\textwidth]{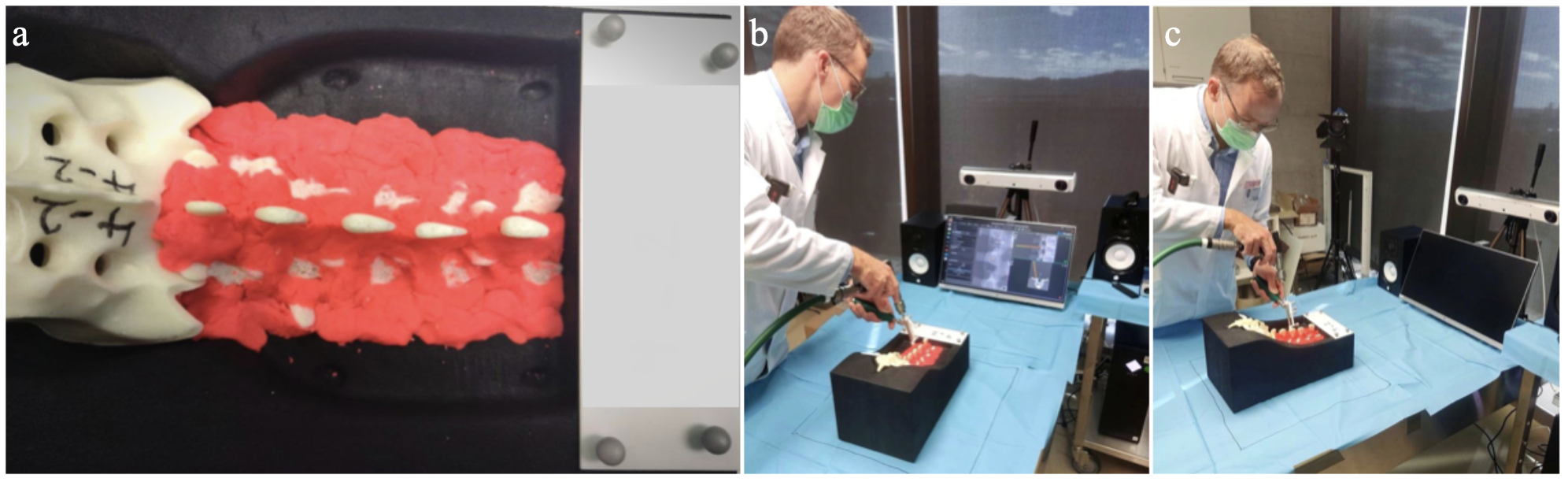}
\end{center}
\caption{The experiment setup, (a) the phantom covered with Play-Doh, (b) task assisted with visualization, (c) task assisted with sonification.}~\label{setup}
\end{figure}

To compare the sonification and the conventional visual navigation methods, we conducted an experiment with 17 orthopedic surgeons, 4 senior experts, and 13 assistant surgeons. In the study, participants performed the pedicle screw placement procedure on phantoms. We used phantom models of the lower lumbar spine (manufactured by Synbone AG, Zizers, Switzerland) consisting of vertebrae L1--L5. The phantoms incorporate facet joints and discs, which create more realistic, intervertebral movement. To simulate the surrounding anatomical landmarks similar to the real surgical environment, we covered the phantoms with Play-Doh to hide the deeper and medial areas around the drilling surface, as shown in Figure~\ref{setup}. Each surgeon drilled 20 pedicle screws on two phantoms with an alternating order between auditory and visual navigations. Our primary measures were the entry point distance error and angular error between the executed and preplanned trajectories. For the procedure with conventional 3D visual navigation, participants performed the four-DOF alignment based on three cross-sectional CT slices from three views. The coronal view visualizes $e_{x}$ and $e_{y}$ on the $P_{entry}$ plane, aligned to the 3D anatomical coordinate system. The axial view visualizes $e_{\phi}$ on the $XY_{a}$ plane. The sagittal view corresponds to $e_{\delta}$ on the $YZ_{a}$ plane (Fig.~\ref{cross-sections}). For the visual model, similar to the sonification model, tracking markers are used to track the drilling sleeve's position relative to a reference marker fixed on the phantom's bed. The real-time processing of the tracking data is performed by the workstation and transferred to the visualization module, which renders the image on a visual display. Figure~\ref{setup} shows the experimental environment.

Starting with a preoperative CT of one of the phantoms, a senior spine surgeon planned 10x lumbar pedicle screws on L1--L5. The preplanned trajectories were aligned to each phantom before starting the trials using a landmark registration method. For the landmark registration, eight points were collected on the most lateral section of each transverse process on L1--L4. L5 was excluded because we observed slight variations among L5 levels in different phantoms; therefore, a higher error for L5 evaluation would be expected.

We used the fusionTrack 500 real-time optical tracking system (Atracsys) and passive infrared markers for tool tracking. The tracking targets on the drill sleeve (3.2~mm, No. 03.614.010, Synapse System) and the phantom's bed were designed with four passive spheres on each. Pivot calibration~\cite{pivot} was performed on the drill sleeve target to transform the tracking coordinates to the center of the drill sleeve's tip. The real-time processing of tracking data (at 50~Hz) was implemented using ImFusionSuite\footnote{ImFusion GmbH, Munich, Germany -- https://www.imfusion.com} software. The generation of sounds was implemented with the SuperCollider3\footnote{https://supercollider.github.io/} software platform for audio synthesis. The communication between the ImFusionSuite and SuperCollider modules was established using the OSC networking protocol~\cite{osc}. Finally, the generated audio signal was sent to a pair of two-way bass reflex studio monitors to be played for the surgeons.

The working area's radius, $r_{EP}$, was set to 20~mm, and the working area's angle, $\theta_{ANG}$ was set to 30\degree. The target zone's thresholds for both alignment phases (EP and AP) were set to 2~mm and 1.5\degree, and the transition zones' thresholds were set to 0.5~mm and 0.375\degree. Choice of these parameters was based on a pilot experiment with an expert spine surgeon, and the optimum values depend on the accuracy of the tracking system, registration, and calibrations. 

Each participant was presented with a short introduction about the method ($\approx$ 5~min). The trials consisted of two phases, a training and an execution phase. In the training phase, the participants were asked to conduct 10 alignment tasks with the aid of sonification, on L1--L5 on both sides of the phantom. In the execution phase, they were asked to conduct the alignment and drilling on two phantoms, resulting in 20 executions on the same vertebrae levels. The executions were divided into four sequences, and each sequence was assisted with either visualization (V) or sonification (S). We randomized the order of the sequences between subjects as $V, S, V, S$ or $S, V, S, V$. Each subject started from either the left or right side of the first phantom and the opposite side of the second phantom, again in a uniformly randomized order. 

The secondary outcome measures were the alignment time and the participants' cognitive. The alignment time is considered the duration between two events, namely, the alignment start and the drilling starting points. This was performed by the trial examiner, pushing a button for each event to record their timestamps. The cognitive load was assessed by asking the participants to respond to a questionnaire, including the NASA Task Load Index (NASA-TLX), a subjective workload assessment measure, and four additional questions provided by the authors. The additional questions were as follows: Q1: Which method helped you better to find the target entry point? Q2: Which method helped you better to find the target angle? Q3: How do you evaluate the overall usability of both systems? Q4: Which navigation feedback method would you like to use in the future?

\section{Evaluation and Results}
\label{results}
\subsection{Evaluation}
We compared the preoperative planned trajectories with the postoperative CT of the drilled phantoms. To detect the exact drilled path, cylindrical graphite sticks with the same diameter as the drill ($3~mm$) were inserted into the phantoms before taking the postoperative CT. The average length of the graphite sticks was $5~cm$. The centers of the first and last disks of each graphite were manually labeled for every drilled screw. We also marked the actual point where the drill had entered into the bone phantom. The actual entry point might be slightly deeper in the bone phantom compared with the planned trajectory because part of the bone surface was removed by surgeons in order to create a flat surface on the pedicle to stabilize the drill sleeve and prevent sliding; a similar procedure is performed during real surgery.

The preoperative and postoperative CT volumes were registered using an image-based registration method. To minimize the movement and possible deformation between vertebrae, we did not remove the Play-Doh before taking the postoperative CT, which caused different appearances in between CT volumes. To resolve this issue, we masked the image-based registration within a 3-mm area around the segmented vertebrae surface. The registration algorithm was manually initialized within its capture range, and a nonlinear optimizer with the LC\textsuperscript{2}~\cite{lc2} similarity metric was used to register the volumes.

\subsection{Results}

\begin{figure}[!ht]
    \subfloat{%
    \includegraphics[clip,width=\textwidth]{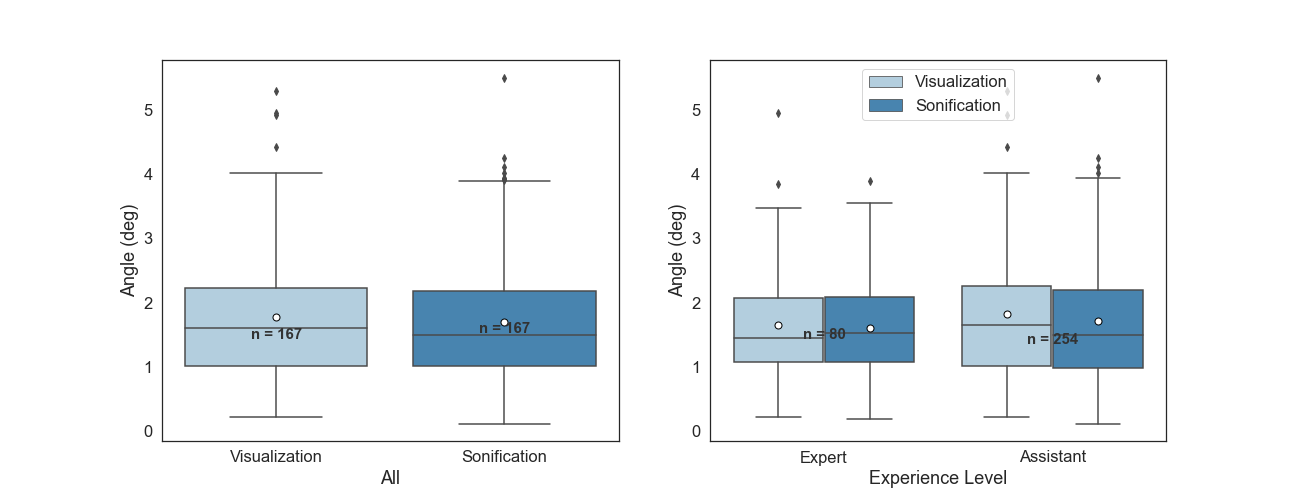}%
    }

    \subfloat{%
    \includegraphics[clip,width=\textwidth]{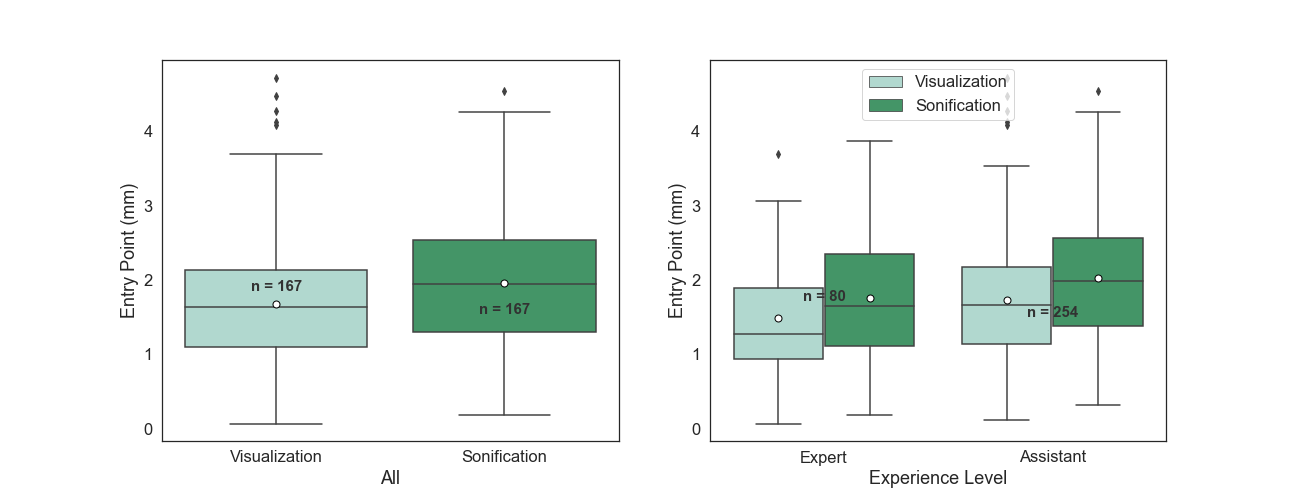}%
    }
    \caption{The mean (CT) error of angle (Top), entry point (Bottom) over expertise groups. The white dots in the middle of box plots represent the mean.}~\label{actual-error}
\end{figure}

\begin{figure}[!ht]
    \begin{center}
        \includegraphics[width=\textwidth]{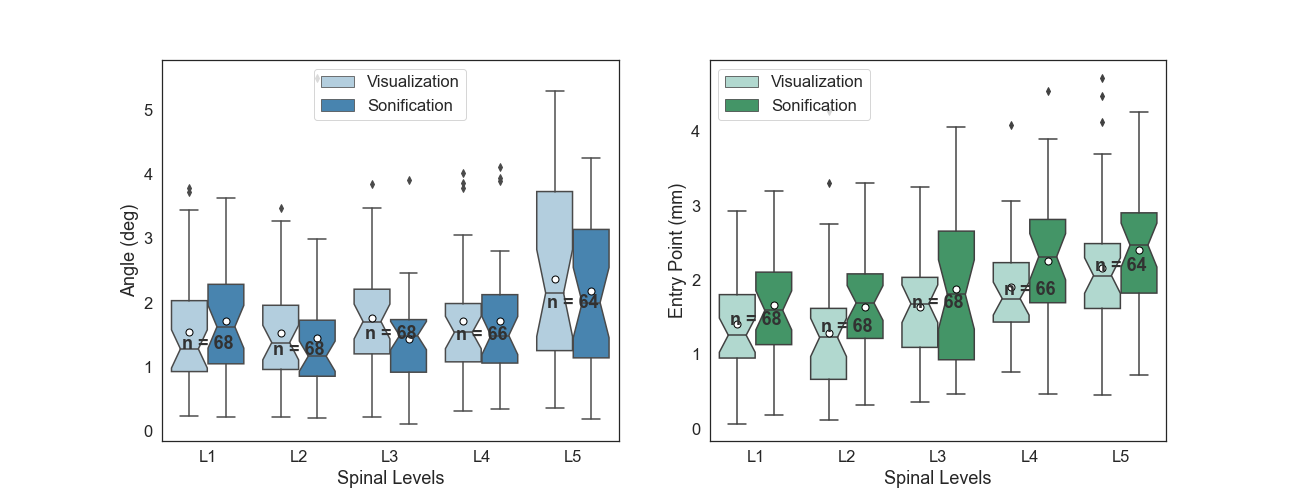}
    \end{center}
    \caption{The mean (CT) error of angle on the left and entry point on the right, per spinal levels. The white dots in the middle of box plots represent the mean.}~\label{levels}
\end{figure}

The results of the post-CT analysis revealed a total mean error of $1.82~mm \pm 0.89~mm$ for the entry point and $1.75\degree~\pm~1.01\degree$ for the angle as deviation from the planned trajectories (CT error, $n~=~336$). Conversely, the system-generated data, which were used to generate both visual and auditory feedback modalities, resulted in a mean error of $0.82~mm~\pm~0.46~mm$ and $0.88\degree~\pm~0.47\degree$ (feedback error, $n~=~323$). We estimated our system error (tracker, registration, and calibration) by subtracting pair-wise samples of the errors (CT and feedback, $n~=~314$) to a mean of $0.98~mm~\pm~0.77~mm$ and $0.82\degree~\pm~0.92\degree$. The mean CT error for visualization ($n~=~167$) was $1.67~mm~\pm~0.87~mm$ and $1.78\degree~ \pm~1.04\degree$ and for sonification ($n~=~167$) $1.96~mm~\pm~0.88~mm$ and $1.69\degree~\pm~0.96\degree$. The details of the error over the expertise groups and the spinal levels are presented in Figures~\ref{actual-error} and \ref{levels} and Table~\ref{mean-error-levels}.

\begin{table}[!ht]
    \caption{The mean error of angle (ANG) and entry point (EP) over the vertebrae levels L1 to L5. The highlighted cells satisfy the safety requirement suggested by Rampersaud et al. ~\cite{accuracy-requirements}}\label{mean-error-levels}
        \centering
        \setlength{\tabcolsep}{7pt} 
        \renewcommand{\arraystretch}{1.5} 
            \begin{tabular}{l l l c c c c c}
                \hline
                Level & & & L1 & L2 & L3 & L4 & L5\\
                \hline
                \hline
                CT error & visualization & EP (mm) & 1.4 & 1.29 & 1.63 & \cellcolor{gray!25}1.91 & \cellcolor{gray!25}2.16\\
                              & & ANG ($\degree$) & 1.55 & 1.52 & 1.76 & \cellcolor{gray!25}1.72 & \cellcolor{gray!25}2.37\\
                 & sonification & EP (mm) & 1.67 & 1.64 & 1.87 & \cellcolor{gray!25}2.26 & \cellcolor{gray!25}2.4\\
                             & & ANG ($\degree$) & 1.71 & 1.45 & 1.43 & \cellcolor{gray!25}1.71 & \cellcolor{gray!25}2.18\\
                \hline
                feedback error & visualization & EP (mm) & 0.51 & \cellcolor{gray!25}0.54 & \cellcolor{gray!25}0.63 & \cellcolor{gray!25}0.71 & \cellcolor{gray!25}0.96\\
                             & & ANG ($\degree$) & 1.02 & \cellcolor{gray!25}0.81 & \cellcolor{gray!25}1.02 & \cellcolor{gray!25}0.84 & \cellcolor{gray!25}1.07\\
                 & sonification & EP (mm) & 0.77 & 0.84 & \cellcolor{gray!25}1.01 & \cellcolor{gray!25}0.99 & \cellcolor{gray!25}1.26\\
                            & & ANG ($\degree$) & 0.88 & 0.78 & \cellcolor{gray!25}0.76 & \cellcolor{gray!25}0.8 & \cellcolor{gray!25}0.82\\
                \hline
            \end{tabular}
\end{table}

\begin{figure}[!ht]
    \begin{center}
        \includegraphics[width=\textwidth]{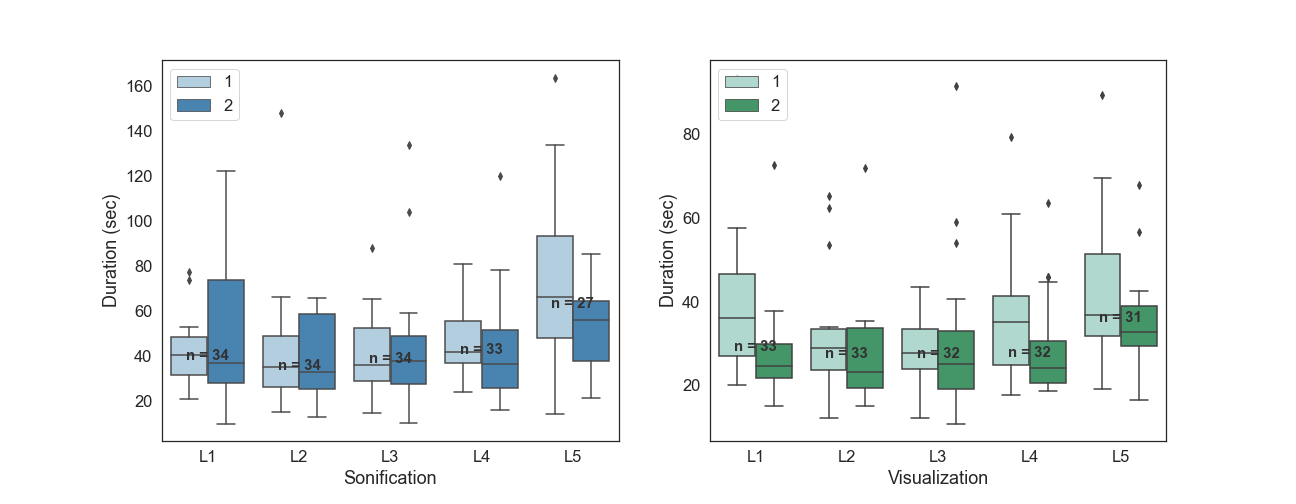}
    \end{center}
    \caption{Alignment time of the 1st. and 2nd. executions over the spinal levels.}~\label{time-levels}
\end{figure}

We expect $< 3$-s error margin in the recording process for the completion time of each alignment. The mean alignment time for visualization was $33.5~s~\pm~16.1~s$ and for sonification $44.1~s~\pm~21.6~s$. The details of the alignment time of the first and second executions on the same level are shown in Figure~\ref{time-levels}.

Fourteen individuals (3 experts and 11 assistant surgeons) returned the questionnaires. We analyzed the NASA-TLX data, using a t-test for independent samples with unequal variances. The results indicated a P-value of 0.59. Therefore, we failed to reject the null hypothesis of having equal means in the samples. Accordingly, we cannot find any significant differences regarding the cognitive load between sonification and visualization. The responses to Q1 and Q2 (which method better helped to find the target entry point and the target angle, respectively) had the same proportions; i.e., 71.4\% voted for visualization, 21.4\% for sonification, and 7.1\% believed there was no difference between both methods. In response to Q3 (overall usability), 42.9\% responded that visualization was more usable than sonification, 35.7\% were of the opinion that both methods were equally usable, 21.4\% believed the sonification method was better than visualization in terms of usability, and no one chose the option ``none of the methods are acceptable". Finally, in response to Q4 (which method would you like to use in the future), the majority of respondents (85.7\%) preferred a system that combines both methods, 7.1\% voted for each of visualization and sonification, and 0\% chose for the option``none of them".

\section{Discussion}

We proposed four-DOF sonification as a novel method for pedicle screw placement, investigating an alternative method toward multisensory assistive technology in the surgical context. The challenge was to design a clinically compatible and accurate system that simultaneously fulfills usability requirements and is competitive with the more conventional visual peer. The results of the comparison study against the state of the art, as demonstrated in Section~\ref{results}, offer clear support of the idea behind this research.

\subsection{Accuracy}
Many clinical and anatomic studies have considered the accuracy of pedicle screw placement as the rate of successful screw placements. A successful screw placement has often been referred as the one fully contained in the pedicle cortex without any degree of perforation. The violating degrees of misplacement have been defined as: $<2~mm$ (Grade A), 2--4~$mm$ (Grade B), and $>4~mm$ (Grade C)~\cite{systematic-pedicle,accuracy-review,accuracy-invivo}. Considering this definition, the accuracy of pedicle screw placement using 2D and 3D visual navigation has been reported at 84.3\% and 95.5\% respectively~\cite{accuracy-review}. To determine whether a particular system will enable the safe performance of the task, we need to specify the safety requirements, as well. The clinical safety requirements are dependent on the type of procedure and the patient's anatomy. The margin of error for a given pedicle is dependent on different factors, such as the size of the screw and the critical dimensions of the pedicle, such as isthmus. Rampersaud et al.~~\cite{accuracy-requirements} proposed a mathematical analysis method for calculating safe margins for the pedicle screw placement task: the maximum entry point and angular error tolerances for L1--L5, given 6.5-mm pedicle screws, are 0.65--3.8~mm and 2.1--12\degree, respectively. The entry point error was defined as the distance between the actual screw insertion point and the ideal starting point for the screw (at the central axis of the pedicle) and the angular error as the angular deviation between the screw trajectory and the ideal trajectory (parallel to the central pedicle axis). For the same pedicle, the error tolerances increase when using a smaller-diameter screw. Similar to this approach, we calculated the error based on the deviations of the actual drilling trajectories from the preplanned targets (the ideal trajectories). 

The overall accuracy of our navigation setup needs to be sufficiently appropriate for the pedicle screw placement task such that we can conduct a valid comparison between the sonification and visualization methods. The accuracies for L4 and L5 in both modalities satisfy the accuracy requirements suggested by Rampersaud et al.~\cite{accuracy-requirements}, as highlighted in Table~\ref{mean-error-levels}. Conversely, the results for L1--L3 do not fully meet these requirements (assuming a 6.5-mm-diameter screw). However, during the actual procedure, the surgeon first drills a guiding hole, with 3-mm diameter in our case, and then inserts a wider screw, with 6.5-mm diameter, which enables the surgeon to manually refine the trajectory based on haptic feedback and the mechanical constraints of the pedicle wall. Therefore, the practical safety thresholds would provide slightly higher tolerance than the suggested thresholds of Rampersaud's. Moreover, as the state-of-the-art navigation method for pedicle screw placement has not yet provided a 100\% success rate, we conclude that the navigation setup has provided an acceptable range of accuracy to compare the sonification and visualization methods. The evaluation of the sonification condition's error ($1.96~mm,~1.69\degree$) indicated a similar accuracy to the visualization condition ($1.67~mm,~1.78\degree$), both demonstrating a better result in comparison with those in~\cite{AR-pedicle1} ($3.35~mm,~2.74\degree$) and ~\cite{AR-pedicle4} ($2.77~mm,~3.38\degree$). Considering the estimated system error ($0.98~mm,~0.73\degree$), which includes registration errors, calibration errors, and tracking data noise, we assume the lower error boundary of $0.65~mm~\text{and}~0.84\degree$ for visualization and $0.96~mm~\text{and}~0.79\degree$ for sonification.

Investigating to what extent both methods have a similar effect, we applied the equivalence test for two independent samples, which is the two one-sided t-test (TOST). TOST works on an equivalence interval (EI) with lower and upper limits $(-\Delta_{L}, \Delta_{U})$, and two composite null hypotheses H0-1: $\Delta~\leq-\Delta_{L}$ and H0-2: $\Delta~\geq~\Delta_{U}$. If both hypothesis tests can statistically be rejected, we can conclude that the difference between sonification and visualization samples, $\Delta$, falls within the EI -- $-\Delta_{L}~<~\Delta~<\Delta_{U}$, which is considered equivalent. The results indicate equivalence of both methods within the EI $\pm~0.46~mm$, $\pm~0.23\degree$ ($P~<~0.05,~n~=~155$). Adding the upper limit of the resultant EI to the sonification error, we can estimate errors of $2.42~mm~\text{and}~1.92\degree$, which is still comparable with the state-of-the-art visual navigation~\cite{AR-pedicle1,AR-pedicle2}. Details of the EI for different expertise groups for actual and feedback errors are presented in Table~\ref{ttost-accuracy}. In general, we observed a larger EI for the entry point compared with the angle, which is because the thresholds for entry point $2~mm$ were set larger, compared with the angle $1.5\degree$. Considering that the threshold mechanism in sonification does not allow the user to obtain any feedback after the threshold level, discussion about accuracy after this level would not be relevant, and the sonification outcome could present a random effect. The thresholds were empirically set during the pilot study as a compromise between accuracy and user's convenience. Having a system design with more accurate tracking, the thresholds can also be reduced, leading to a more accurate result for sonification. 

\begin{table}[!ht]
    \caption{The least EI of the TOST to reach the CI $P < 0.05$, over the expertise groups, including both CT and feedback errors for entry point (EP) and angle (ANG).}\label{ttost-accuracy}
        \centering
        \setlength{\tabcolsep}{6.5pt} 
        \renewcommand{\arraystretch}{1.5} 
            \begin{tabular}{|c|c|c|c|c|c|c|c|c|c|c|c|}
                \hline
                Group & \multicolumn{2}{c|}{All $(n=155)$} & \multicolumn{2}{c|}{Experts $(n=38)$} & \multicolumn{2}{c|}{Assistants $(n=117)$}\\
                \hline
                Error & EP & ANG & EP & ANG & EP & ANG\\
                \hline
                CT & $\pm 0.46~mm$ & $\pm 0.3\degree$ & $\pm 0.5~mm$ & $\pm 0.28\degree$ & $\pm 0.5~mm$ & $\pm 0.23\degree$\\
                \hline
                Feedback & $\pm 0.39~mm$ & $\pm 0.24\degree$ & $\pm 0.45~mm$ & $\pm 0.23\degree$ & $\pm 0.41~mm$ & $\pm 0.2\degree$\\
                \hline
            \end{tabular}
\end{table}

\subsection{Learning Curve}
To determine the training effect in both expertise groups, we compared completion time and errors, as functions of performance, on two consecutive executions on the same vertebrae level. To confirm the learnability, we have to determine whether the mean duration of the second execution decreased compared with that of the first execution, without significant decrease in accuracy. Hence, we conducted a paired samples t-test on the alignment time of both executions. The P-values for each expertise group are shown in Table~\ref{ttest-time-attempt}. Moreover, to determine any decrease in accuracy, we performed a paired samples t-test ($\alpha~<~0.05$) on both entry point and angle errors, which failed to reject the null hypothesis in terms of equal means for both samples. The mean differences between both executions are shown in Table~\ref{diff-2errors-attempt}.

The mean difference between both errors (Table~\ref{diff-2errors-attempt}) and the results of the t-test indicate that the accuracy remained consistent during both executions for all expertise groups. As presented in Table~\ref{ttest-time-attempt}, the experts demonstrated a significant decrease in time for visualization and sonification at the confidence level of 99\% and 90\%, respectively. We can observe this pattern in the less experienced group only for visualization ($P~<~0.05$). Considering the error consistency and the observed time patterns, we can conclude high learnability for the both modalities for the experts; however, this was demonstrated by the assistants only for the visualization condition. 

Our interpretation is that the experience level of the expert group enabled them to focus more on learning the untrained auditory navigation method. Conversely, the assistant surgeons required more of their cognitive processing capacity for executing the screw drilling task, and therefore, they had a lower learning rate in the sonification condition. Because the visual navigation was more familiar for both groups, they could improve their speed on this modality. However, further research is required to accurately evaluate the effect of training for the sonification method and for developing a full picture of its learning curve.

\begin{table}[!ht]
    \caption{The P-values of the paired samples t-test on the alignment time of the 1st. and 2nd. executions.}\label{ttest-time-attempt}
        \centering
        \setlength{\tabcolsep}{7.5pt} 
        \renewcommand{\arraystretch}{1.5} 
            \begin{tabular}{|c|c|c|c|}
                \hline
                Group & All ($n=77$) & Experts ($n=19$) & Assistants ($n=58$)\\
                \hline
                Visualization & 0.004 & 0.01 & 0.02\\
                \hline
                Sonification & 0.34 & 0.1 & 0.59\\
                \hline
            \end{tabular}
\end{table}

\begin{table}[!ht]
    \caption{The mean of differences for entry point (CT) error, angle (CT) error, and alignment time between the 1st. and 2nd. executions for each expertise group.}\label{diff-2errors-attempt}
        \centering
        \setlength{\tabcolsep}{7.5pt} 
        \renewcommand{\arraystretch}{1.5} 
            \begin{tabular}{|c|c|c|c|}
                \hline
                Group & All & Experts & Assistants\\
                \hline
                \multirow{2}{*}{Visualization} & $5.96~sec,$ & $4.97~sec,$ & $6.29~sec,$\\
                & $0.15~mm, -0.11\degree$  & $0.072~mm, 0.07\degree$  & $0.17~mm, -0.17\degree$\\
                \hline
                \multirow{2}{*}{Sonification} & $1.58~sec,$ & $7.07~sec,$ & $-0.26~sec,$\\
                & $-0.09~mm, -0.04\degree$ & $0.02~mm, 0.02\degree$ & $-0.12~mm, -0.05\degree$\\
                \hline
            \end{tabular}
\end{table}

\subsection{Multisensory Processing and Research Outlook}
Although the plurality of the questionnaire respondents preferred visualization (Q1--Q3), the absolute majority (85\%) imagined that a desirable future system would combine the advantages of both modalities. Such responses from the field's experts are absolutely in accordance with the equivalent results of the accuracy, NASA-TLX assessments, and principles of multisensory perception. Multisensory solutions result in increased performance and recall, in particular, in intense and complex sensory scenarios~\cite{Burg-Multisensory,Ngo-Multisensory,Barutchu-Multisensory}. Research questions that could be asked in future studies include, e.g., \emph{To what extent can each modality convey complex information accurately? When is it better, or preferable, for the perceptual modalities to be presented in a complementary fashion, and in which situations do they have to provide redundant contextual information? How do our decisional resources respond to each perceptual cue?} Future research in computer-assisted surgery can focus on investigating the possible answers to such fundamental questions in the application field of surgical navigation, as the foundation is well established in cognitive science~\cite{Ernst-Multisensory,Helbig-Multisensory}. 

Further, previous sonification research~\cite{sonificationnavigation1,sonificationnavigation2,sonificationnavigation3,sonificationnavigation4,sonificationnavigation5,sonificationnavigation6,sonificationnavigation7,sonificationnavigation8} has investigated different sonification strategies for navigation, providing the preliminary basis for further research. Future studies should take the cons and pros of sonification paradigms into account. For instance, spatial sonification as an intuitive and natural method with a high learnability rate is suitable for orientation tasks~\cite{sonificationnavigation5,sonificationnavigation8}; however, we cannot not disregard its limitations with respect to resolution~\cite{psychoac}. Additionally, spatial sonification may cause localization anomalies such as front--back confusion, a vague distance and elevation perception, and orientation errors~\cite{spatial1,spatial2}. On the other hand, monaural sonification as a candidate approach provides efficiency~\cite{psychoac,jnd} and flexibility in design. Nonetheless, it requires a more prolonged learning phase, which can also depend on the design concept and parameterization. 

Four-DOF sonification is a useful tool for scenarios with complex dimensionality and accuracy challenges, as demanded by surgical applications. We divided multiple dimensions into subsets and controlled switching between them using a threshold mechanism. This idea is expandable to contexts with higher dimensionality. Nonetheless, issues such as intuitiveness and the learning process have to be considered. Finally, we should consider that monaural sonification is a rather new approach, evolving in terms of dimensionality and interaction design. Even though the presented study exhibited promising results, future research should investigate the effect of enhanced learning phases on performance.

\section{Conclusions}
In this article, we have outlined the problem of complex data perception in high-intensity environments, such as the operating room, and highlighted the importance of multisensory processing and development of creative solutions to overcome the information overload issue. To investigate the effects of multisensory processing, we conducted a study with 17 medical professionals in a lab environment using a spinal bone phantom and compared two different techniques for surgical navigation assistance. We proposed the four-DOF sonification method -- as a stand-alone audio-based solution -- for navigating pedicle screw placement in spinal fusion surgery and compared the method with state-of-the-art visual navigation. Four-DOF sonification demonstrates statistically equivalent performance compared with the visual navigation, satisfying the clinical requirements of pedicle screw placement. The novel design concept of the method supports the idea for accurate sonification of high-dimensional data within a complex interactive task scenario. This study is the first step toward enhancing our understanding of perceptual multisensory processing in the surgical context.

\bibliographystyle{splncs04}
\bibliography{mybibliography}

\end{document}